\documentstyle[eqsecnum,aps,manuscript]{revtex}
\def\b{\begin{equation}}
\def\e{\end{equation}}
\def\be{\begin{eqnarray}}
\def\ee{\end{eqnarray}}
\def\bm{\begin{mathletters}}
\def\em{\end{mathletters}}
\def\o{\over}
\def\a{\alpha}
\def\t{\triangle}

\begin{document}

\draft

\title{SUPERSYMMETRIC $SO(10)$ WITH $SU(2)_L\times SU(2)_R\times 
SU(4)_C$ INTERMEDIATE GAUGE SYMMETRY}

\author{M.K.Parida\footnote{E-mail:mparida@nehus.ren.nic.in}\footnote{ICTP 
Associate}\footnote{Permanent
Address:
Physics department,North--Eastern Hill University
Shillong 793 022, India.}}

\address{High Energy Physics Group\\ 
International Centre for Theoretical Physics\\
I-34100 Trieste, Italy\\ and}

\author{S.Dey and B.Purkayastha}

\address{Physics Department, North--Eastern Hill University\\ 
Shillong 793 022, India}
\date{\today}
\maketitle

\begin{abstract}

In  recent analyses the existence of $SU(2)_L\times 
SU(2)_R\times SU(4)_C(g_{2L}\neq g_{2R})$ intermediate gauge 
symmetry has been ruled out in SUSY $SO(10)$ model at one--loop level,
although the left-right symmetric intermediate gauge group has been
shown to exist with certain light scalar superfields near 1TeV. 
We show how the asymmetric gauge group  is allowed 
with an intermediate scale $M_I=10^{10}-10^{13}$ GeV by including 
two--loop and threshold effects ,but without  any light degrees 
of freedom.

\end{abstract}

\pacs{12.10.Dm, 12.60.Jv}

\narrowtext

\section{INTRODUCTION}

One of the major motivations in following SUSY $SO(10)$ grand unified 
theory is its potentiality to explain fermion masses and mixings 
[1] and ,in particular,neutrino masses over a wide 
range of values  via simple see-saw mechnism[2], or with specific
textures in mass matrices[3].The observed cosmological baryon asymmetry of
the universe can be also explained  by triggering baryogenesis via
leptogenesis,if right-handed Majorana neutinos are superheavy[4].   
Apart from the interesting possibility that a massive 
$\nu_\tau\,(m_{\nu_\tau}\simeq 2-10{\rm eV})$ is a promising candidate 
for hot dark matter of the universe,experimental hints  on solar neutrino 
deficit  could be explained
through matter enhanced MSW effects[5] via see-saw prediction of
left-handed neutrino masses provided, the
right-handed neutrinos  have  masses in the range of 
$M_N\simeq 10^{10}-10^{13}$ GeV[6]. This might
be realized in single step breaking of SUSY $SO(10)$ if the Yukawa coupling 
of $\overline{\underline{126}}$ to matter spinors is adjusted to be
small ,or via dim-4 nonrenormalizable couplings between matter multiplets
and Higgs fields belonging  to $\overline{\underline{16}}$.Prospects of
solar neutrino oscillation in supergrand desert model with right-handed 
Majorana neutrino masses at intermediate scales have been discussed
in ref.[6].But the most attractive possibility is to relate 
$M_N$ to an intermediate scale$(M_I)$ corresponding to the
spontaneous symmetry breaking of the intermediate 
gauge group such as $SU(2)_L\times SU(2)_R\times U(1)_{B-L}\times 
SU(3)_C\,(\equiv G_{2213})$ or  
$SU(2)_L\times SU(2)_R\times SU(4)_C\,(\equiv G_{224})$[7] 
without having the necessity to adjust the Majorana type $\nu_R$-Yukawa
coupling to very small values.Such an intermediate scale 
also solves the strong CP problem by Peccei-Quinn mechanism[8]. 
Recently, although the existance of $G_{2213}$ intermediate gauge 
symmetry with decoupled parity and $SU(2)_R$ -breaking $(g_{2L}\ne g_{2R})$ 
 [9] has been established in a series of papers [10-12], the intermediate
gauge symmetry $G_{224} (g_{2L}\ne g_{2R})$ has been ruled out [11].
However more recently, it has been shown that the 
$G_{224P} (g_{2L}= g_{2R})$ intermediate gauge symmetry 
with unbroken left-right discrete symmetry $(\equiv {\rm Parity}(P))$ 
can survive down to the intermediate scale of $M_I\simeq 10^{12}-10^{13}$ 
GeV provided the model permits light Higgs supermultiplets
near the TeV scale [13]. $G_{224}$ is the maximal subgroup of $SO(10)$ 
which contains the quark-lepton unification of Pati-Salam [7] and has 
one gauge-coupling constant less as compared to $G_{2213}$. 
All the gauge couplings of $G_{224}$ are determined 
through the CERN-LEP data and the intermediate-scale matching
conditions,
\[{\a_{4C}(M_I)=\a_{3C}(M_I)}\nonumber\] 
\b{1\o \a_Y(M_I)}={3\o 5}{1\o\a_{2R}(M_I)}+{2\o 5}{1\o \a_{4C}(M_I)}\e
The see-saw formulas ,for neutrino  masses where up-quark masses appear
instead of the Dirac-neutrino masses [2] ,emerge more naturally at the
intermediate scale due to the presence of quark-lepton symmetry
in $G_{224}$.
   
\par
The purpose of this paper is to show how the $G_{224} (g_{2L}\ne g_{2R})$
intermediate gauge symmetry ,with parity broken at the GUT scale ,is
allowed to survive naturally down to the 
desired intermediate scale by the 
inclusion of two--loop [14] and threshold effects [15] in SUSY $SO(10)$.
To achive SO(10) breaking to $G_{224}$,we use the Higgs 
representation $\underline {54}$ in addition to 
$\underline{210}$[16] and break $G_{224}$ by 
$\underline{16}\oplus\overline{\underline{16}}$,instead of $\underline
{126}\oplus\overline{\underline{126}}$ ,to avoid large one-loop
contributions of
the triplets in the latter upsetting solutions to renormalization group
equations(RGEs).
\par
In Section II we derive analytical formulas for   mass scales. In
Section III threshold effects and solutions to mass scales 
are obtained using  method of effective mass parameters of ref.[15].  A
brief summary with conclusion is provided in Section IV.
                     
\section{ANALYTIC FORMULAS FOR MASS SCALES} 

In this section we derive analytic formulas for the unification mass $M_U$ 
and the intermediate scale $M_I$ including one--loop, two--loop [14] and 
threshold contributions [15]. We consider the following model using the 
mechanism of decoupling parity and $SU(2)_R$ -breakings [9],
\FL\[ SO(10)\times{\rm SUSY}\buildrel M_U\over \longrightarrow 
G_{224}\times {\rm SUSY}\buildrel M_I\over \longrightarrow\nonumber\]
\FR\b G_{213}\times 
{\rm SUSY}\buildrel M_Z\over \longrightarrow U(1)_{em}\times SU(3)_C\e
where  $G_{213}$ is the standard gauge symmetry 
$SU(2)_L\times U(1)_Y\times SU(3)_C$.
In the first step of (2.1), the combined effect  of
$\underline{54}$ and $\underline{210}$ ,containing $G_{224}$ -singlets,
break D-Parity and SO(10)  without breaking $G_{224}$.
In the second step ,we use two sets of $\underline{16}\oplus
\underline{\overline{16}}$ . The right--handed doublets 
$(1,2,4)\oplus (1,2,\overline{4})$ contained in 
($\underline{16}\oplus \underline{\overline{16}}$) are kept
lighter having masses near $M_I$ whereas the left--handed counterparts 
$(2,1,4)\oplus (2,1,\overline{4})$ acquire masses near $M_U$. In the 
third step of (2.1) we use a representation $\underline{10}$ containing 
the $u$-- and the $d$--type Higgs doublets to break the symmetry to
$U(1)_{em}\times SU(3)_C$.
The renormalization group equations in the presence of the two gauge 
symmetries $G_{213}$ and $G_{224}$ can be written as
\FL\[{1\o\a_i(M_Z)}={1\o\a_i(M_I)}+{a_i\o 2\pi}\ln{M_I\o M_Z}+
{1\o 4\pi}P_i-\t_i\nonumber\]
\FR\b{i=1Y,2L,3C}\e
\FL\[{1\o\a_i(M_I)}={1\o\a_i(M_U)}+{a'_i\o 2\pi}\ln{M_U\o M_I}+
{1\o 4\pi}P'_i-\t'_i\nonumber\]
\FR\b{i=2L,2R,4C}\e
where the second (third) terms in the R.H.S. of (2.2)-(2.3) represent 
one--loop (two--loop) contributions with
\[ P_i=\sum_jB_{ij}\ln{\a_j(M_I)\o\a_j(M_Z)}\nonumber\]
\[ P'_i=\sum_jB'_{ij}\ln{\a_j(M_U)\o\a_j(M_I)}\nonumber\]
\b B_{ij}={b_{ij}\o a_j}\;,\;B'_{ij}={b'_{ij}\o a'_j}\e
Here $a_j(b_{ij})$ and $a'_j(b'_{ij})$ are the one--loop (two--loop) 
$\beta$--function coefficients in the two mass ranges and their values 
are given below. The terms $\t_i$ and $\t'_i$ in the R.H.S. of equations 
(2.2)-(2.3) represent threshold effects at $\mu=M_Z,M_I$ and $M_U$ with 
\[\t_i=\t^Z_i+\t^I_i\nonumber\] 
The function $\t_i^Z$ includes
threshold effects at $\mu=M_Z$ due to the top quark-Yukawa coupling, 
and masses of Higgs scalars and superpartners in the SUSY standard model 
different from $M_Z$, but $\t^I_i$ represents threshold effects due to the 
Higgs scalars and superpartners having masses near $M_I$. $\t'_i$ takes
into account threshold effects due to Higgs scalars and their 
superpartners having masses near $M_U$. Such scalars are contained in 
\underline {54},
\underline{210}, $\underline{16}\oplus \underline{\overline{16}}$ and 
\underline{10}. Although one set of 
$\underline{16}\oplus \underline{\overline{16}}\subset SO(10)$
is sufficient to break the intermediate gauge symmetry to the standard SUSY
gauge theory, we investigate the effects of two sets of such spinorial 
representations to achive desired one-loop solution. Expressions for
$\t_i$ and $\t'_i$ are given in Sec.III. 
Using suitable combinations of gauge couplings and equations (2.2)-(2.3), 
we obtain the following analytic formulaes for mass scales, $M_I$ and $M_U$,
\FL\[\ln{M_I\o M_Z}={L_SA_U-L_\theta B_U\o D}+{K_\theta B_U-J_\theta 
A_U\o D}+\nonumber\]
\FR\b {J_\t A_U-K_\t B_U\o D}\e
\FL\[\ln{M_U\o M_Z}={L_\theta B_I-L_SA_I\o D}+{J_\theta A_I-K_\theta 
B_I\o D}+\nonumber\]
\FR\b{K_\t B_I-J_\t A_I\o D}\e
where \[D=A_UB_I-A_IB_U\nonumber\]
\[L_\theta ={16\pi\o\a (M_Z)}\left(\;{3\o 8}-\sin^2\theta_W\;\right)
\nonumber\]  
\b L_S={16\pi\o\a (M_Z)}\left(\;{3\o 8}-{\a (M_Z)\o\a_S(M_Z)}\;\right)\e
\[A_U=3a'_{2R}+2a'_{4C}-5a'_{2L}\nonumber\]
\[A_I=5a_{1Y}-5a_{2L}-3a'_{2R}-2a'_{4C}+5a'_{2L}\nonumber\]
\[B_U=3a'_{2R}+3a'_{2L}-6a'_{4C}\nonumber\] 
\[B_I=5a_{1Y}+3a_{2L}-8a_{3C}-3a'_{2R}-3a'_{2L}+6a'_{4C}\nonumber\] 
\[J_\theta ={1\o 2}\left(\;3P'_{2R}+3P'_{2L}-6P'_{4C}+5P_{1Y}+
3P_{2L}-8P_{3C}\;\right)\nonumber\] 
\[K_\theta ={1\o 2}\left(\;3P'_{2R}+2P'_{4C}-5P'_{2L}+5P_{1Y}-
5P_{2L}\;\right)\nonumber\] 
\FL\[J_\t=\nonumber\]
\FR\[ 2\pi\left(\;3\t'_{2R}+3\t'_{2L}-6\t'_{4C}+5\t_{1Y}+
3\t_{2L}-8\t_{3C}\;\right)\nonumber\] 
\b K_\t=2\pi\left(\;3\t'_{2R}-5\t'_{2L}+2\t'_{4C}+5\t_{1Y}-
5\t_{2L}\;\right)\e 
An attractive feature of the analytic formulas given in (2.5)--(2.6) is 
that, in the R.H.S., contributions due to every loop order or threshold 
effects are seperated out. For example, the first, the second and the 
third terms in the R.H.S. of (2.5)--(2.6) represent, analytically 
one--loop, two--loop and threshold corrections, respectively. 
\par
The one-- and two--loop $\beta$--function coefficients for the MSSM [14] 
are
\[ a_i=\left(\;\begin{array}{c}1\\ {33/5}\\ -3\end{array}\;\right)
\nonumber\]
\FL\[ B_{ij}={b_{ij}\o a_j}=\left(\;\begin{array}{ccc}25&3/11&-8\\ 
27/5&199/65&-88/15\\ 9&1/3&-14/3\end{array}\;\right)\;,\nonumber\]
\FR\b i,j=2L,1Y,3C\e
In the presence of $G_{224}\times SUSY$ intermediate symmetry in the 
mass range $\mu=M_I-M_U$, we use the contributions from the Higgs 
scalars and their superpartners contained in the representations 
\underline{10} and two sets of $\underline{16}\oplus
\underline{\overline{16}}$ of $SO(10)$. The components which have masses 
near $M_I$ are the $G_{224}$ 
-submultiplets $\phi\,(2,2,1)$ and two sets of $(1,2,4)\oplus 
(1,2,\overline{4})$. Other components of the $SO(10)$ representations                          
such as $(1,1,6)\subset \underline{10}$ and two sets of $(2,1,4)\oplus 
(2,1,\overline{4})\subset 2(\underline{16}\oplus\underline{\overline{16}})$    
have masses near $M_U$. Following the standard procedure and including 
contributions of gauge bosons, fermions, Higgs scalars, and their 
superpartners, the one--loop and two--loop coefficients for the $G_{224}$ 
symmetry are computed as,
\[ a'_i=\left(\;\begin{array}{c}1\\ 9\\ -2\end{array}\;\right)\nonumber\]
\b B'_{ij}={b'_{ij}\o a'_j}=\left(\;
\begin{array}{ccc}25&1/3&-45/2\\ 3&13/3&-30\\ 9&4/3&-25/4\end{array}\;
\right)\;,\;i,j=2L,2R,4C\e
Using the values of the coefficients from eqs. (2.9)--(2.10) in 
(2.7)--(2.8), we obtain
\b A_U=18\;,\;A_I=10\;,\;B_U=42\;,\;B_I=18\;,\;D=-96\e
In the next section we derive expressions for threshold effects and 
present solutions to the mass scales.

\section{LOWERING THE INTERMEDIATE SCALE BY THRESHOLD EFFECTS}

 Including only one--loop and two--loop contributions, 
the expressions for $M_I$ and $M_U$ are given by the first and the second 
terms, respectively, in the R.H.S. of eqs.(2.5)--(2.6). Using eqs.(2.7) and 
(2.11), the one--loop and two--loop contributions are
\[\left(\;{L_SA_U-L_\theta B_U\o D}\;\right)_{\rm one-loop}=
{\pi\o\a}\left(\;{3\o 2}+3{\a\o\a_s}-7\sin^2\theta_W\;\right)\nonumber\]
\[\left(\;{L_\theta B_I-L_SA_I\o D}\;\right)_{\rm one-loop}=
{\pi\o\a}\left(\;3sin^2\theta_W-{1\o 2}-{5\o 3}{\a\o\a_s}\;\right)
\nonumber\]
\FL\[\left(\;{K_\theta B_U-J_\theta A_U\o D}\;\right)_{\rm two-loop}=
\nonumber\]
\FR\[{1\o 8}\left(\;11P'_{2L}-3P'_{2R}-8P'_{4C}+11P_{2L}-5P_{1Y}-6P_{3C}\;
\right)\nonumber\]
\FL\[\left(\;{J_\theta A_I-K_\theta B_I\o D}\;\right)_{\rm two-loop}=
\nonumber\]
\FR\b{1\o 48}\left(\;6P'_{2R}-30P'_{2L}+24P'_{4C}+15P_{2L}-35P_{1Y}+
20P_{3C}\;\right)\e
Using (3.1) in the first two terms on the R.H.S. of (2.5)--(2.6), we 
obtain the expressions for the mass scales upto two--loop order as

\widetext

\FL\[\ln{M_I\o M_Z}={\pi\o\a}\left(\;{3\o 2}+3{\a\o\a_s}-
7\sin^2\theta_W\;\right)+\nonumber\]
\FR\bm\be {1\o 8}\left(\;11P'_{2L}-3P'_{2R}-8P'_{4C}+
11P_{2L}-5P_{1Y}-6P_{3C}\;\right)\ee
\FL\[\ln{M_U\o M_Z}={\pi\o\a}\left(\;3sin^2\theta_W-{1\o 2}-
{5\o 3}{\a\o\a_s}\;\right)+\nonumber\]
\FR\be {1\o 48}\left(\;6P'_{2R}-30P'_{2L}+
24P'_{4C}+15P_{2L}-35P_{1Y}+20P_{3C}\;\right)\ee\em

\narrowtext

For numerical analysis we use the following input parameters [17]
\[\a^{-1}(M_Z)=128.9\pm 0.1\nonumber\]
\[\a_{3C}=0.119\pm 0.004\nonumber\]
\[\sin^2\theta_W=0.2315\pm 0.0003\nonumber\]
\b{M_Z=91.18 GeV}\e
While solving for $M_I$ and $M_U$, using eqs. (3.2a)-(3.2b) by including 
only one--loop contributions and ignoring two--loop effects, we obtain 
$M_I=8.30\times 10^{14}$ GeV, $M_U=1.01\times 10^{17}$ GeV. .
But, as one important result of 
this paper, we show that when threshold effects  near $M_U$, $M_I$ and  
$M_Z$[15] are included, along with one--loop and two--loop effects, 
the model yields  $M_I$ substantially lower than 
$M_U$,which itself is consistent with  string unification scale.The
threshold effects at $M_Z$ have been already computed[15]. For
calculating these effects at $ M_I$ we also follow the method of effective
mass parameters[15].At first, 
we seperate $J_\t$ and $K_\t$ into three different parts,
\[J_\t=J^U_\t+J^I_\t+J^Z_\t\nonumber\] 
\b K_\t=K^U_\t+K^I_\t+K^Z_\t\e 
where
\[J^U_\t=2\pi\left(\;3\t'_{2L}+3\t'_{2R}-6\t'_{4C}\;\right)\nonumber\]   
\[K^U_\t=2\pi\left(\;3\t'_{2R}+2\t'_{4C}-5\t'_{2L}\;\right)\nonumber\]
\[J^i_\t=2\pi\left(\;5\t^i_{1Y}+3\t^i_{2L}-8\t^i_{3C}\;\right)\nonumber\]   
\b K^i_\t=2\pi\left(\;5\t^i_{1Y}-5\t^i_{2L}\;\right)\;,\;i=I,Z\e
The expression for $\t^Z_i$ is given by[15], 
\FL\[ \t^Z_i=\t^{\rm conversion}_i+\t^{\rm Yukawa}_i+\t^{\rm SUSY}_i\;,
\nonumber\]
\FR\b i=1Y,2L,3C\e
\b\t^{\rm conversion}_i=-{C_2(G_i)\o 12\pi}\e
where $C_2(G_i)$ is the quadratic Casimier operator for the adjoint 
representation, with
\[ C_2(G_i)=N[0]\; for\; G_i=SU(N)[U(1)]\nonumber\]
In eq.(3.6) 
\[ \t^{\rm Yukawa}_i=b^{\rm top}_i{h^2_t\o 16\pi^2}t\nonumber\]
\b b^{\rm top}_i=\left(\;\begin{array}{c}26/5\\ 6\\ 4\end{array}\;\right)\e
In the present case \[ t={1\o 2\pi}\ln{M_I\o M_Z}\nonumber\]
In terms of effective mass parameters $(M_i,\,i=1,2,3)$ near 
$M_Z$--threshold [15], the superpartner contributions in (3.6) are
\[ \t^{\rm SUSY}_{1Y}={5\o 4\pi}\ln{M_1\o M_Z}\nonumber\]  
\[ \t^{\rm SUSY}_{2L}={25\o 12\pi}\ln{M_2\o M_Z}\nonumber\]  
\b \t^{\rm SUSY}_{3C}={2\o \pi}\ln{M_3\o M_Z}\e
For the sake of convenience we use $M_1=M_2=M_3=6M_Z$[15].

\par

The superheavy components contained in the two sets of 
$\underline{16}\oplus \underline{\overline{16}}$
which have masses near $M_I$ are given in Table I. The corresponding 
threshold effects can be expressed in terms of the effective mass 
parameters $(M'_i)$ as
\[ \t^I_i={b'_i\o 2\pi}\ln{M'_i\o M_I}\;,\;i=1Y,2L,3C\nonumber\]
\b b'_i=\sum_\a b'^{(\a)}_i\e
where $\a$ includes Higgs scalar components and their superpartners 
near $M_I$. The superheavy components in the representations
under $G_{224}$,cotained in $\underline{54}$, $\underline {210}$,
$(\underline{16}
\oplus\underline{\overline{16}})$,and $\underline{10}$,  
which have masses near the GUT scale are shown in Table II. 
The expression for the threshold effects $\t'_i$ is given by
\[ \t'_i={b'^{\prime}_i\o 2\pi}\ln{M''_i\o M_U}\;,\;i=2L,2R,4C
\nonumber\]
\b b'^{\prime}_i=\sum_\a b''^{(\a)}_i\e
where $\a$ includes Higgs scalar components and their superpartners 
near $M_U$, and $M''_i$ are the effective mass parameters.
Using eqs.(3.10)--(3.11) in eqs.(3.4)--(3.5) we obtain
\FL\[ J_\t=150\ln{M''_{2R}\o M_U}+174\ln{M''_{2L}\o M_U}-
342\ln{M''_{4C}\o M_U}\nonumber\]
\[ +25\ln{M'_{1Y}\o M_I}-28\ln{M'_{3C}\o M_I}+
{25\o 2}\ln{M_1\o M_Z}+{25\o 2}\ln{M_2\o M_Z}\nonumber\]
\FR\b -32\ln{M_3\o M_Z}+{3\o 4\pi^2}\ln{M_U\o M_Z}+3\e
\FL\[ K_\t=150\ln{M''_{2R}\o M_U}+114\ln{M''_{4C}\o M_U}-
290\ln{M''_{2L}\o M_U}\nonumber\]
\[ +25\ln{M'_{1Y}\o M_I}+{25\o 2}\ln{M_1\o M_Z}-
{125\o 6}\ln{M_2\o M_Z}\nonumber\]
\FR\b -{1\o 4\pi^2}\ln{M_U\o M_Z}+{5\o 3}\e
Using eq.(3.12)--(3.13) in eqs.(2.5)--(2.8) we then obtain the formulas 
for threshold effects on $M_I$ and $M_U$ in terms of effective mass 
parameters $M'_i$ and $M''_i$.
\FL\[ \t\ln{M_I\o M_Z}\equiv {\left(\;J_\t A_U-K_\t B_U\;\right)\o D}
\nonumber\]
\[ ={75\o 2}\ln{M''_{2R}\o M_U}+144\ln{M''_{4C}\o M_U}-
{319\o 2}\ln{M''_{2L}\o M_U}\nonumber\]
\FR\b +{25\o 4}\ln{M'_{1Y}\o M_I}+{21\o 4}\ln{M'_{3C}\o M_I}-4.89\e
\FL\[ \t\ln{M_U\o M_Z}\equiv {\left(\;K_\t B_I-J_\t A_I\;\right)\o D}
\nonumber\]
\[ =-{25\o 2}\ln{M''_{2R}\o M_U}-57\ln{M''_{4C}\o M_U}+
{145\o 2}\ln{M''_{2L}\o M_U}\nonumber\]
\FR\b -{25\o 12}\ln{M'_{1Y}\o M_I}-{35\o 12}\ln{M'_{3C}\o M_I}+1.93\e
The last two numbers in eqs.(3.14)--(3.15) denote threshold contributions 
at $\mu =M_Z$ corresponding to the choice $M_1=M_2=M_3=6M_Z$. 
In our analysis the effective mass parameters $M'_i$ 
or $M''_i$ are taken to vary between $1/5-5$ times the 
relevant scale of symmetry breaking i.e. $M_I$ or $M_U$. For example when 
$M'_{1Y}=M'_{3C}=2.3M_I\;,\;M''_{2L}=3M_U\;,\;M''_{2R}=2M_U\;,\;
M''_{4C}=3M_U$
we obtain the values of $M_I\simeq 1.75\times 10^{10}$ GeV and 
$M_U\simeq 7.9\times 10^{17}$ GeV. But, when $M'_{1Y}=M'_{3C}=
3M_I,\;M''_{2L}=3M_U,\;M''_{2R}=2M_U,\;M''_{4C}=3M_U$, 
the values of $M_I$ and $M_U$ are found to be $3.73\times 10^{11}$ 
GeV, and $2.09\times 10^{17}$ GeV, respectively.
Different results on intermediate scale $M_I$ and unification mass 
$M_U$ which are obtained as solutions of RGEs including 
threshold effects as a function of effective mass parameters  are 
presented in Table III. We find $M_I\simeq 10^{10}-10^{13}$ GeV
for quite reasonable choices of the mass parameters.It is interesting
to note that some of the GUT scales are close to the Plank-scale or the 
string unification scale.The solutions given here are by no means
exhaustive,but indicate that the intrmediate scale can be achieved
in a natural way via threshold effects by following the method of 
effective mass parameters[15].  

\section{SUMMARY AND CONCLUSION}

While investigating the possibility of $G_{224}$
intermediate gauge symmetry in $SO(10)$, we avoided the 
representations $\underline{126}\oplus \underline{\overline{126}}$ for 
intermediate symmetry breaking and generating Majorana neutrino masses 
because of their well known difficulties against ariving at 
acceptable values of $M_I\ll M_U$ [11]. As 
the mechanism of generating Majorana neutrino masses are now well 
known via the representations $\underline{16}\oplus 
\underline{\overline{16}}$ and through couplings with $SO(10)$-singlet
field in the superpotential[12],we 
have utilised two sets of them and, in addition, the representations 
$\underline{210}$ ,$\underline{54}$ and $\underline{10}$ needed for
spontaneous symmetry 
breakings at the GUT and the electroweak scales, respectively. We have 
found that at the two--loop level,  when threshold effects due to superheavy 
components contained in these representations are included, the RGEs 
permit $G_{224}$ -breaking intermediate
scales $M_I\simeq 10^{10}-10^{13}$ GeV with high unification scales, 
$M_U\simeq 10^{17}-10^{18}$ GeV,for certain allowed solutions. The
generation of right--handed 
Majorana neutrino masses and the implementation of seesaw mechanism 
is  carried out by the introduction of $SO(10)$ -singlet  following 
the mechanism of Lee and Mohapatra [12] through purely renormalisable 
interactions. We thus conclude, in contrast to earlier observations 
[11], that $SU(2)_L\times SU(2)_R\times SU(4)_C (g_{2L}\neq g_{2R})$
is allowed as an intermediate gauge symmetry 
in supersymmetric $SO(10)$ model in a natural manner. Even the use of 
a number of light Higgs supermultiplets at the intermediate scale [11,12]
is not needed to achieve the intermediate scale.Using quadratic or
linear see-saw formulas and renormalisation effects[18],it is possible to
obtain neutrino masses necessary for ${\nu}_{\tau}$ as a hot dark matter
candidate and solution to the solar neutrino puzzle by MSW mechanism,
in these models. 

\acknowledgements
One of us (M.K.P) thanks the Professor Goran Senjanovic for
useful discussion and to Professor Randjbar-Daemi, the High
Energy Group ,International Centre for Theoretical
Physics,Trieste,Italy for Hospitality.Two of us (M.K.P.
and B.K.P.) acknowledge financial assistance from the Department of
Science and Technology, New Delhi  through  the research project No.
SP/S2/K--09/91.

\begin{table}
\caption{The heavy Higgs content of the $SO(10)$ model with $G_{224}$
intermediate symmetry. The $G_{213}$ submultiplets 
become massive when $G_{224}$ is broken. In the extreme right column
of the Table are threshold contributions $b'_i$ of the different
multiplets.}
\begin{tabular}{ccc}
{$SO(10)$ representation}&{$G_{213}$ multiplet}&
{$b'_{2L},\,b'_{1Y},\,b'_{3C}$}\\ 
\tableline\\
{$\underline{16}$}&{$(1,{1\o 3},\overline{3})$}&
{$(0,{1\o 5},{1\o 2})$}\\ \\
{$\overline{16}$}&(1,-1,1)&{$(0,{3\o 5},0)$}\\
&(1,0,1)&(0,0,0)\\ 
&{$(1,-{1\o 3},3)$}&{$(0,{1\o 5},{1\o 2})$}\\
&{$(1,-{2\o 3},3)$}&{$(0,{4\o 5},{1\o 2})$}\\ \\
16&(1,1,1)&{$(0,{3\o 5},0)$}\\
&(1,0,1)&(0,0,0)\\
&{$(1,{1\o 3},\overline{3})$}&{$(0,{1\o 5},{1\o 2})$}\\
&{$(1,-{2\o 3},\overline{3})$}&{$(0,{4\o 5},{1\o 2})$}\\ \\
{$\overline{16}$}&(1,-1,1)&{$(0,{3\o 5},0)$}\\
&(1,0,1)&(0,0,0)\\
&{$(1,-{1\o 3},3)$}&{$(0,{1\o 5},{1\o 2})$}\\
&{$(1,-{2\o 3},3)$}&{$(0,{4\o 5},{1\o 2})$}
\end{tabular}
\end{table}

\begin{table}
\caption{Same as Table I, but here the $G_{224}$ submultiplets acquire
masses when $SO(10)$ is broken. Also listed, in the
extreme right column of the Table, are the threshold contributions 
$b''_i$ of different multiplets.}
\begin{tabular}{ccc}\\ 
{$SO(10)$ representation}&{$G_{224}$ multiplet}&
{$b''_{2L},\,b''_{2R},\,b''_{4C}$}\\ 
\tableline\\
210&(2,2,10)&(10,10,12)\\
&{$(2,2,\overline{10})$}&(10,10,12)\\
&(1,1,15)&(0,0,4)\\
&(1,3,15)&(0,30,12)\\
&(3,1,15)&(30,0,12)\\ \\
54&(3,3,1)&(2,2,0)\\
&(1,1,20)&(0,0,6)\\
&(2,2,6)&(6,6,4)\\ \\
16&(2,1,4)&(2,0,1)\\ \\
{$\overline{16}$}&{$(2,1,\overline{4})$}&(2,0,1)\\ \\
10&(1,1,6)&(0,0,1)
\end{tabular}
\end{table}

\begin{table}
\caption{Predictions on mass scales $M_I$ and $M_U$ including threshold 
effect with effective mass parameters.}
\begin{tabular}{ccccccc}
{$M'_{1Y}$}&{$M'_{3C}$}&{$M''_{2L}$}&{$M''_{2R}$}&
{$M''_{3C}$}&{$M_I$ (GeV)}&{$M_U$ (GeV)}\\ 
\tableline\\
{${1\o 2}M_I$}&{${1\o 2}M_I$}&{$2M_U$}&{$2M_U$}&{$2M_U$}&
{$4.3\times 10^{10}$}&{$3.03\times 10^{18}$}\\
{$2.3M_I$}&{$2.3M_I$}&{$3M_U$}&{$2M_U$}&{$3M_U$}&
{$1.75\times 10^{10}$}&{$7.90\times 10^{17}$}\\
{$3M_I$}&{$3M_I$}&{$3M_U$}&{$2M_U$}&{$3M_U$}&
{$3.73\times 10^{11}$}&{$2.09\times 10^{17}$}\\
{$3.5M_I$}&{$3.5M_I$}&{$3M_U$}&{$2M_U$}&{$3M_U$}&
{$2.19\times 10^{12}$}&{$9.68\times 10^{16}$}\\
{$4M_I$}&{$4M_I$}&{$3M_U$}&{$2M_U$}&{$3M_U$}&
{$1.02\times 10^{13}$ }&{$4.96\times 10^{16}$}

\end{tabular}
\end{table}

\end{document}